\documentclass[fleqn,usenatbib]{mnras}

\usepackage{newtxtext,newtxmath}

\usepackage[T1]{fontenc}

\DeclareRobustCommand{\VAN}[3]{#2}
\let\VANthebibliography\thebibliography
\def\thebibliography{\DeclareRobustCommand{\VAN}[3]{##3}\VANthebibliography}

\usepackage{graphicx}	
\usepackage{amsmath}	
\usepackage{natbib}
\usepackage{hyperref}
\usepackage{esint}

\title[Sound waves driven by AGN jets]{Production Efficiencies of Sound Waves in the Intracluster Medium Driven by AGN Jets}

\author[]{
Shiang-Chih Wang,$^{1}$
H.-Y.~Karen Yang$^{1,2,3}$\thanks{E-mail: hyang@phys.nthu.edu.tw}
\\
$^{1}$Institute of Astronomy, National Tsing Hua University, Hsinchu 30013, Taiwan\\
$^{2}$Center for Informatics and Computation in Astronomy, National Tsing Hua University, Hsinchu 30013, Taiwan\\
$^{3}$Physics Division, National Center for Theoretical Sciences, Taipei 10617, Taiwan
}

\date{Accepted XXX. Received YYY; in original form ZZZ}

\pubyear{2022}

\begin{document}
\label{firstpage}
\pagerange{\pageref{firstpage}--\pageref{lastpage}}
\maketitle

\begin{abstract}
Feedback from active galactic nuclei (AGN) is believed to be the most promising solution to the cooling flow problem in cool-core clusters. Dissipation of sound waves is considered as one of the possible heating mechanisms; however, its relative contribution to heating remains unclear. To estimate the energy budget for heating, we perform 3D hydrodynamic simulations of AGN jet injections in a Perseus-like cluster and quantify the amount of energy stored in the forms of weak shocks and waves. We find that, for a single jet injection with typical parameters in cool-core clusters, $\sim9\%$ of the total jet energy is stored in compressional waves (including both shocks and waves). However, due to the destructive effects among randomly phased waves as well as the dissipation of shock energies, in our simulations including self-regulated AGN feedback, no more than $3\%$ of the total injected energy goes into compressional waves. We further separate the energy contribution from shocks and waves and find that, for a single outburst, the shocks can only contribute to $\sim20-30\%$ of the total compressional energy in the inner radii and quickly dissipate away. In the self-regulated case where shocks are repeatedly generated, shocks completely dominate over sound waves in the inner region and can still provide $\sim40-50\%$ of the total compressional energy at outer radii. Our results suggest that the production of sound waves is not as efficient as what was previously found, and thus sound wave dissipation may be a subdominant source of heating in cool-core clusters.
\end{abstract}

\begin{keywords}
galaxies: active -- galaxies: clusters: intracluster medium -- hydrodynamics -- methods: numerical
\end{keywords}



\section{Introduction}
X-ray observations by Chandra and XMM-Newton reveal that the radiative cooling time of some galaxy clusters is much shorter than the Hubble time. These clusters, the so-called cool-core (CC) clusters, are expected to drive strong cooling flows at rates $100-1000$ $\mathrm{M_{\odot}}$ yr$^{-1}$ assuming the isobaric cooling-flow model in the absence of heat sources. However, observationally there is lack of gas below $\sim1$ keV (\citealt{Peterson03}). In addition, the observed star formation rates are typically 1-2 orders of magnitude lower than values predicted by the classical cooling flow model (\citealt{Hoffer12}; \citealt{Donahue15}). These discrepancies are referred to as "the cooling flow problem" (\citealt{Fabian77}; \citealt{Fabian94}). Thanks to numerous efforts over the past two decades, it has become widely accepted that active galactic nucleus (AGN) feedback is the primary mechanism that could provide heat to the intracluster medium (ICM) and prevent the cluster cores from catastrophic cooling (\citealt{McNamara12}). However, because of the huge separation scales between the clusters and the accretion disks of the central supermassive black holes (SMBHs), the detailed processes of SMBH feeding and feedback are still a major unresolved problem in astrophysics.        

To better understand the AGN feedback processes, numerical simulations with increasing levels of complexity have been developed. Some of the works have focused on the evolution of an initially static, buoyantly rising bubble \citep[e.g.,][]{Churazov01, Robinson04, Reynolds05, Dong09}, or the detailed interactions between the a jet-inflated bubble and the ambient gas \citep[e.g.,][]{Vernaleo06, Guo08, Guo18, Yang19}. Other works used simulations of self-regulated AGN feedback and demonstrated that a global thermal stability could be established between radiative cooling and the heat provided by the AGN jets \citep[e.g.,][]{Sijacki07, Gaspari11, Yang16b, Li17, Prasad17}. These efforts have allowed us to have a closer look at how the AGN energy is converted into heat and is distributed throughout the core.  

In terms of the heating mechanisms of AGN jets, a lot of different physical processes have been proposed and have been extensively studied using numerical simulations. Possible candidates include direct mixing between the bubbles and the ICM \citep{Hillel16, Yang16b, Hillel20}, shock heating \citep{Li17}, turbulent dissipation (\citealt{Zhuravleva14}), sloshing \citep{Zuhone12}, thermal conduction \citep{Zakamska03, Voigt04, Yang16a}, and cosmic rays (CRs) \citep{Guo08, Pfrommer13, R17, Yang19, Kempski20}.  

In addition to the above, another heating mechanism that is often invoked is the dissipation of sound waves, which is the main focus on this work. Sound waves in the ICM have caught attention since 2003, when the ripple structures were first detected in the Perseus cluster (\citealt{Fabian03}). Based on this observational work, numerical simulations showed that viscous dissipation of sound waves can efficiently heat the ICM and balance radiative cooling (\citealt{Ruszkowski04a}; \citealt{Ruszkowski04b}). The propagation of sound waves have also been studied using analytical models \citep{Fabian05, Fujita05, Fabian17}. More recently, sound wave propagation with an improved treatment considering the ion and electron temperature perturbations separately has also been investigated (\citealt{Zweibel17b}). 


In recent years, there have been several attempts to measure the amount of energy stored in the form of sound waves in the ICM, both theoretically and observationally. Thanks to the improvement of numerical techniques, high-resolution hydrodynamic simulations become viable for modeling the process of sound wave production. \cite{Tang17} showed that, with energy injections from a point source under spherical symmetry, no more than $12\%$ of the injected energy can be stored in sound waves. \cite{Bambic19} followed up the previous work and considered an episode of AGN injection via a pair of bipolar jets assuming axisymmetry. They carried out a detailed scan of parameters and found that more than $25\%$ of the total injected energy can be converted into sound waves in optimal cases, suggesting that sound wave dissipation may be a significant source of ICM heating. However, recent observational works based on X-ray hardness ratios proposed an alternative interpretation that the ripples in the Perseus cluster may be associated with stratified turbulence instead of sound waves \citep{Zhuravleva14b}. Data analyses on the Perseus cluster showed that $\sim 50\%$ of the surface brightness fluctuations at the location of the ripples come from isobaric perturbations (e.g., turbulence), whereas the adiabatic perturbations (e.g., shocks and sound waves) only have a minor ($\sim 5\%$) contribution \citep{Zhuravleva16}. The discrepancies between the numerical works and the observational results motivate further studies to clarify the importance of sound waves in CC clusters. 


The goal of our work is to continue the investigation of sound wave production by AGN jets and quantify the energy budgets available for sound wave heating using three-dimensional (3D) hydrodynamic simulations. While some of the previous simulations \citep{Tang17, Bambic19} have focused on the energy partitions for a single AGN outburst, in realistic CC clusters multiple AGN injections are required to balance radiative cooling and establish global thermal equilibrium. To estimate the production efficiencies of sound waves in more realistic CC environments, we carry out two types of simulations in this work: one with a single jet injection with conditions suitable for the Perseus cluster, and another set of simulations including self-regulated AGN feedback with varying feedback efficiencies. We measure the fractions of jet energy converted into compressional waves (including weak shocks and sound waves) following the method of \cite{Bambic19}. We further separate the energy contributions from weak shocks and sound waves and evaluate their relative importance.


The structure of this paper is as follows. In \S\ref{Sec:method}, we describe the setup of the simulations and the methods used to compute energy stored in compressional waves. In \S\ref{Sec:Results}, we present the results for the single-injection case (\S\ref{Sec:single_AGN}) and for the self-regulated feedback simulations (\S\ref{Sec:self_regulated}). In \S\ref{Sec:discussion}, we compare our results with  previous works (\S\ref{Sec:comparison}) and discuss the limitations of our simulations (\S\ref{Sec:limitation}). The conclusions are summarized in \S\ref{Sec:conclusion}.

\section{Methods} \label{Sec:method}
We perform 3D hydrodynamic simulations including bipolar jet outflows within a Perseus-like cluster using the adaptive-mesh-refinement (AMR) code FLASH (\citealt{Flash}). The bipolar jets are assumed to be injected along the $z$-axis from the SMBH at the center of the simulation grid. The simulations solve the standard hydrodynamic equations:
\begin{equation}
\frac{\partial\rho}{\partial t}+\nabla\cdot(\rho \boldsymbol{v})=0
\end{equation}
\begin{equation}
\frac{\partial (\rho \boldsymbol{v})}{\partial t}+\nabla\cdot(\rho \boldsymbol{v} \boldsymbol{v})+\nabla P=\rho \boldsymbol{g}
\end{equation}
\begin{equation}
\frac{\partial E_{\mathrm{tot}}}{\partial t}+\nabla\cdot\left[ \left(E_{\mathrm{tot}} +P\right)\boldsymbol{v}\right]=\rho \boldsymbol{v}\cdot\boldsymbol{g}-n_{\rm e}^2\Lambda(T)
\end{equation}
where $\rho$ and $\boldsymbol{v}$ are the density and velocity of the gas, $P$ is the thermal pressure, $\boldsymbol{g}$ is the gravitational acceleration, and $E_{\mathrm{tot}}=e+\rho\left |\boldsymbol{v} \right |^2/2$ is the total energy density including the internal energy density $e$ and kinetic energy density. Radiative cooling of the ICM is included as the $n_{\rm e}^2\Lambda(T)$ term, where $n_{\rm e}=\rho/\mu_{\rm e}m_{\rm p}$ is the electron number density and $\Lambda(T)$ is the cooling function. Because the dissipation of sound waves due to thermal conduction and viscosity is dependent on plasma properties of the ICM, which remains poorly understood, we choose not to include viscosity and thermal conduction in the current simulations. This means that, in this work, we are only concerning the total energy budget that is available for sound wave dissipation, rather than the actual energy dissipated, similar to the approach adopted by \cite{Tang17} and \cite{Bambic19}. We recognize that the above transport processes are important mechanisms for dissipating sound waves, and therefore the amount of sound wave energy estimated in this work is likely an upper limit because of this omission (see detailed discussion in \ref{Sec:discussion}).  

The analyses of wave energies were performed on two types of simulations: one with a single jet injection, and another set of simulations with self-regulated AGN feedback. The simulation setup for the single outburst case is identical to the KIN run in \cite{Yang19}, to which we refer the readers for more details. The jet energy is dominated by kinetic energy. With a simulation box size of 500 kpc and a maximum refinement level up to 8, the minimum cell size is 0.49 kpc in the simulation. In order to capture the small-scale structures of the perturbations, the inner 100 kpc region from the cluster center is forced to be refined to the highest resolution. The power of the jet is $5\times 10^{45}$ erg $\rm s^{-1}$ over a duration of 10 Myr. This choice of jet parameters corresponds to the average jet activity as found in previous self-regulated simulations of the same initial conditions \citep{Yang16b}, and hence this simulation is meant to study a representative case in a Perseus-like cluster.  

The initial conditions for our second set of simulations are essentially identical to the setups in \cite{Yang16b}, in which radiative cooling and self-regulated AGN feedback are included. We summarize the important ingredients in the methods and highlight the differences; more details can be found in \cite{Yang16b}. In these simulations, cold gas is formed due to local thermal instabilities, and the jet injections are coupled with the mass accretion rates of the central SMBH with an efficiency parameter, establishing a quasi-steady ICM atmosphere in global thermal equilibrium. The main difference between the current work and \cite{Yang16b} is that we do not drop out the cold gas ($T<5\times10^5$ K) in the simulations. This is motivated by a recent work \citep{Wang21}, which shows that cold gas could be an important driver of turbulence in the ICM on kpc scales. In order to capture the dynamical effects of cold gas and, in particular, the generation of waves and turbulence, the cold gas is retained in the current simulations.  

Following previous works modeling cold gas accretion \citep{Li17,Yang16b}, the SMBH accretion rate can be written as:
\begin{equation}
\dot{M}_{\mathrm{BH}}=\frac{M_{\mathrm{cold}}}{t_{\mathrm{acc}}}
\end{equation}
where $M_{\mathrm{cold}}$ is the mass of cold gas within a sphere of radius 2 kpc around the central black hole, and the typical accretion timescale $t_{\mathrm{acc}}=5$ Myr is assumed to be comparable to the free-fall time near the cluster center. The injection rate of mass, momentum, and energy by the AGN jets can then be computed as \citep{Yang12}:
\begin{equation}
\begin{split}
&\dot{M}_{\mathrm{jet}}=\eta\dot{M}_{\mathrm{BH}}\\
&\dot{p}_{\mathrm{jet}}=\sqrt{2\epsilon\eta}\dot{M}_{\mathrm{BH}}c\\
&\dot{E}_{\mathrm{jet}}=\epsilon\dot{M}_{\mathrm{BH}}c^2
\end{split}
\end{equation}
where $\eta=1$ is the mass loading factor, $c$ is the speed of light, and $\epsilon$ is the feedback efficiency. The above terms are injected via a cylindrical jet nozzle with radius 2 kpc and half height of 2 kpc. In our fiducial case, we adopt $\epsilon=10^{-3}$; we also considered two additional cases with $\epsilon=10^{-2}$ and $\epsilon=10^{-4}$ for comparisons. The maximum refinement level in this simulation is 7, which corresponds to the minimum resolution element of 0.98 kpc. A small-angle jet precession is assumed to stir up the ambient medium to mimic turbulence seen in real clusters. 


The calculation of wave energies is an important part of this paper. We follow the method of \cite{Bambic19} to measure the amount of energy stored in sound waves by analyzing gas properties in each simulation output file. The method is briefly summarized below. The perturbed gas pressure, density and velocity can be defined as $\delta P=P-P_0$, $\delta \rho=\rho-\rho_0$ and $\delta \boldsymbol{v}=\boldsymbol{v}$, where $P_0$ and $\rho_0$ are the initial pressure and density. The conservation law of acoustic waves can be written as (see \cite{Bambic19} for a derivation):
\begin{align}
&\frac{\partial u_{\mathrm{w}}}{\partial t}+\nabla\cdot \boldsymbol{F}_{\rm w}=\boldsymbol{F}_{\rm w}\cdot \left ( \frac{\nabla P_0}{\gamma P_0}-\frac{\nabla \rho_0}{\rho_0}\right)\label{eq:conservation_law}\\
&u_{\mathrm{w}}=\frac{\delta P^2}{2\gamma P_0}+\frac{1}{2}\rho_0 \left | \delta \boldsymbol{v} \right |^2\\
&\boldsymbol{F}_{\rm w}=\delta P\delta \boldsymbol{v}
\end{align}
where $\boldsymbol{F}_{\rm w}$ is the flux density and $u_{\mathrm{w}}$ is the energy density of acoustic waves, which is equal to the sum of its internal energy density and kinetic energy density. In this expression, the total acoustic energy in a given volume can be estimated as $E_{\mathrm{w}}=\iiint u_{\mathrm{w}} dV$. The right-hand side of Eq.\ \ref{eq:conservation_law} vanishes if the waves are propagating in an isothermal medium in hydrostatic equilibrium. In more realistic situations as well as in our simulation setup, this term is not strictly zero because the ICM in the cool cores is not isothermal. However, we explicitly evaluate this term in our simulations, and we verified that this term is negligible ($\sim 1 \%$) compared to other terms in the equation throughout the simulations. Therefore, this term is neglected in the subsequent analyses. The acoustic power crossing a surface $S$ at radius $r$ can then be computed by integrating Eq.\ \ref{eq:conservation_law} using Gauss's theorem:
\begin{equation}
P_{\mathrm{w}}(r)=\oint \boldsymbol{F}_{\rm w}\cdot \boldsymbol{\hat{n}}\ {\rm d}S,\label{eq:wave_power}
\end{equation}
where d$S$ is the area element of the surface $S$ and $\boldsymbol{\hat{n}}$ is the unit vector normal to $S$. 

Note that when the above method is applied to the simulation data, the wave energies estimated would include not only acoustic/sound waves (linear perturbations with $\delta P/P_0 \ll 1$) but also shock waves (nonlinear discontinuities with non-negligible $\delta P/P_0$; see Figure \ref{fig:delta_profile} and the discussion in Appendix C of \cite{Bambic19}). To avoid confusion, we hereafter refer to the total wave energies estimated using Eq.\ \ref{eq:wave_power} as compressional wave energies instead of acoustic wave energies. While the dissipation of shocks are handled in the FLASH code by default, the sound waves do not dissipate if the transport processes are not explicitly modeled\footnote{Numerical viscosity might cause some dissipation of sound wave energy, though the effect should be insignificant in our simulations (see discussion in \S~\ref{Sec:Results}).}. Because of their different dissipation physics, in some of our later analyses we separate the contributions from shocks and sound waves to the total energy of compressional waves. We identify the shock waves by considering negative velocity divergence and pressure jumps. We show results using two pressure-jump thresholds for selecting shocks: $\Delta P/P>0.1$ and $\Delta P/P>0.2$, where $\Delta P$ is the jump in gas pressure across neighboring cells. The above thresholds correspond to Mach numbers of $\mathcal{M}\sim 1.04$ and $\mathcal{M}\sim 1.07$, respectively, for an ideal gas with adiabatic index of $\gamma=5/3$ \citep{Yang16b}. These thresholds are chosen as they could more clearly pick out visible shock structures (see Figures \ref{fig:delta_profile} and \ref{fig:P_grad}). Some additional criteria are used in other literature to identify shock waves (\citealt{Ryu03}; \citealt{Schaal15}; \citealt{Dubois19}). Thus, our estimations of shock energies are likely to be upper limits. Also, as we will show in \S~\ref{Sec:Results}, the main conclusions are not sensitive to the chosen values. 

\section{Results} \label{Sec:Results}
\subsection{Single AGN outburst}\label{Sec:single_AGN}
\begin{figure*}
	\includegraphics[width=0.8\textwidth]{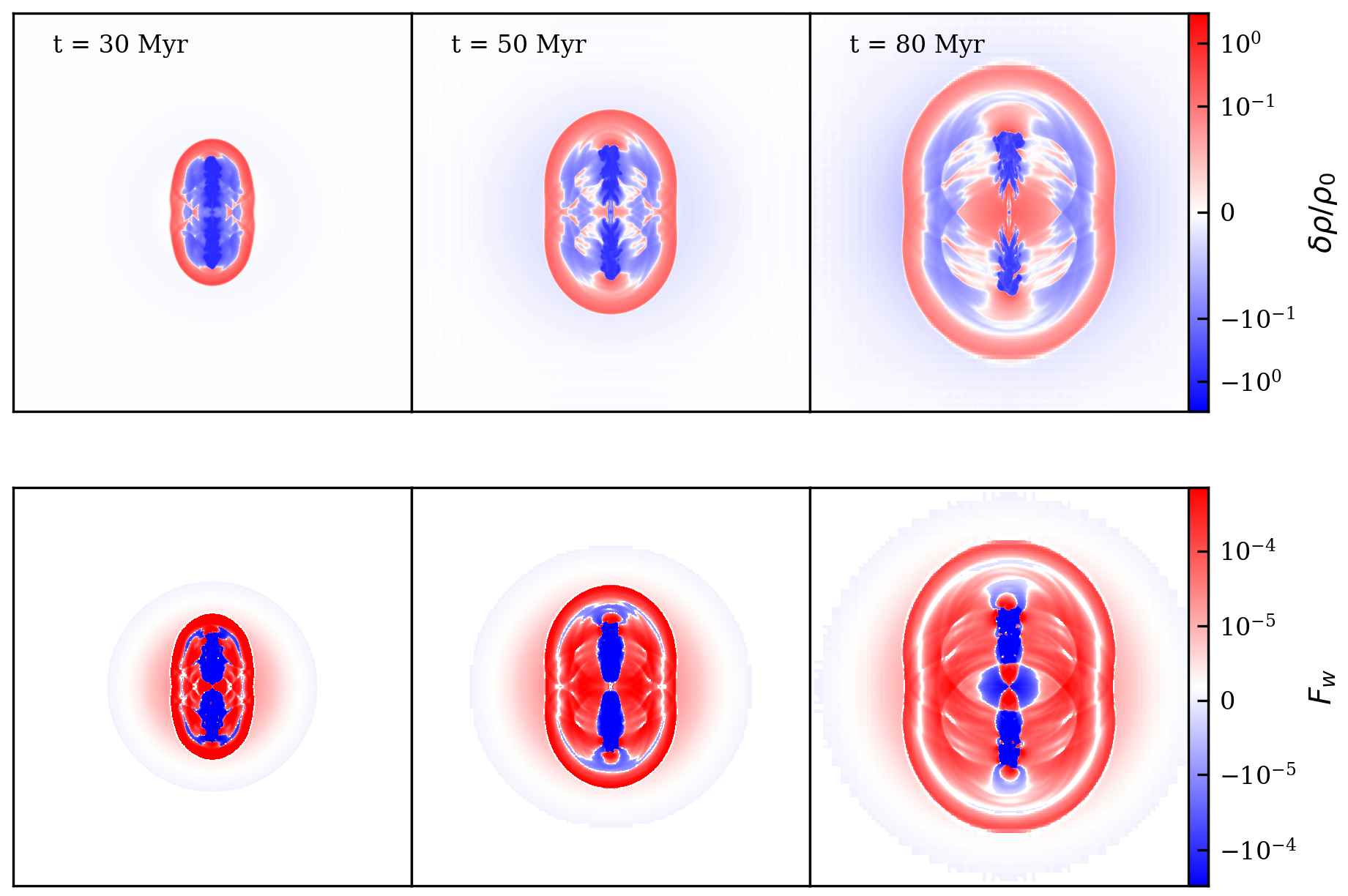}
	\centering
    \caption{Slices of perturbed density $\delta\rho$ (normalized to $\rho_0$; top row) and compressional wave flux density $\boldsymbol{F}_{\rm w}$ (in units of erg s$^{-1}$ cm$^{-2}$; bottom row) within the cluster core at $t = 30, 50$, and 80 Myr for the single-jet simulation. The physical scale of each panel is 350 kpc by 350 kpc. The bottom row shows the evolution of the compressional waves which propagate outward (red) and inward (blue).}
    \label{fig:acoustic_flux_density}
\end{figure*}

In this section, we present results from the simulation of a single jet injection. The evolution of gas densities is shown in the top row of Figure \ref{fig:acoustic_flux_density}. After the initial injection period (10 Myr), the jets are turned off and we follow the subsequent evolution of the ICM. Due to the AGN outburst, the ICM quickly expand with supersonic speeds, forming an outward-propagating shock front. Coincident with the jet fluid are the low-density cavities/bubbles, which are in rough pressure equilibrium with the surrounding ICM and rise buoyantly away from the cluster center. Since our simulation does not include mechanisms that could help suppress the fluid instabilities such as magnetic fields and viscosity, the bubbles are disrupted due to Rayleigh-Taylor (RT) and Kelvin-Helmholtz (KH) instabilities as they rise. The instabilities then drive turbulent motions and sound waves on the scale of the bubbles, which can later interfere and form sound waves on larger scales. Previous simulations demonstrated that this process can allow over $25\%$ of the jet energy to be transferred into compressional waves for optimal jet parameters \citep{Bambic19}. Here we show that, with typical jet parameters and setup suitable for the Perseus cluster, no more than $9\%$ of the total jet energy goes into compressional waves. This fraction is consistent with the previous work when the jet parameters are properly rescaled (see \S~\ref{Sec:comparison}), though it is lower than their optimal estimation.

\begin{figure*}
	\includegraphics[width=0.7\textwidth]{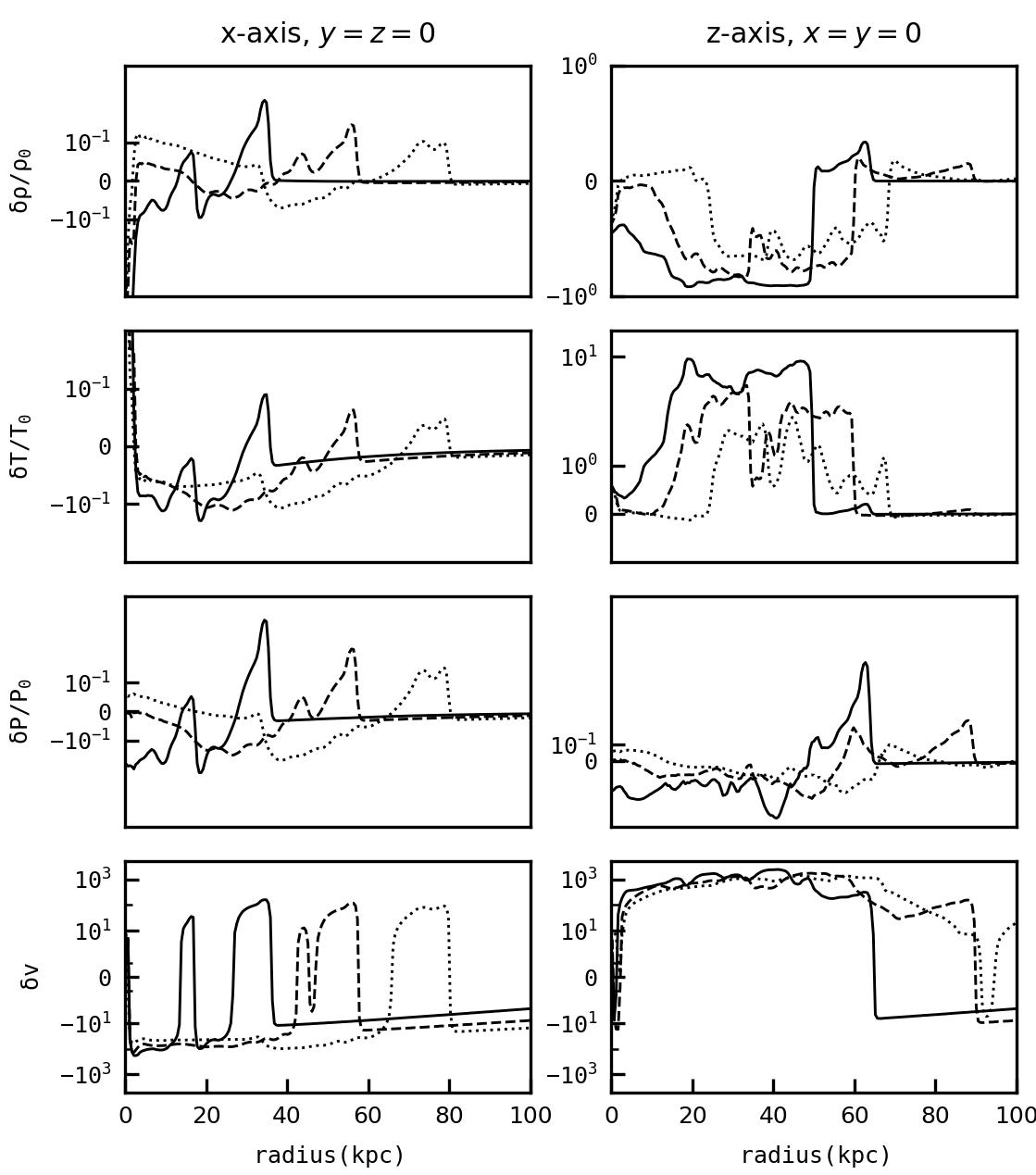}
	\centering
    \caption{Gas profiles for the single-outburst simulation. Panels from top to bottom indicate  profiles for the change of density, temperature, pressure and  radial velocity (in units of km s$^{-1}$) of the gas with respect to their initial values. The profiles are evaluated along the $x$-axis (perpendicular to the jet axis; left column) and along the $z$-axis (along the jet axis; right column) at $t = 30$ (solid line), $t = 50$ (dashed line), and $t = 80$ (dotted line) Myr.}
    \label{fig:delta_profile}
\end{figure*}


    %
\subsubsection{Energy in compressional waves}\label{Sec:acoustic}
Figure \ref{fig:acoustic_flux_density} (bottom row) shows the flux density of compressional waves at three different simulation times, $t = 30$, $50$, and $70$ Myr. Positive values (red colors) refer to outward-propagating waves, and negative values (blue colors) correspond to inward-propagating waves. The outermost elliptical region in the figure with large compressional wave fluxes corresponds to the forward shock that is produced by the AGN outburst. The discontinuities of gas properties at the locations of the shocks could be clearly seen from the horizontal and vertical profiles in Figure \ref{fig:delta_profile}. As time goes on, the shocks dissipate their energies and become sound waves at later times. The AGN jet-inflated bubbles, characterized by low gas densities and high temperatures, can also be seen from the gas profiles. In the vicinity of the bubbles, the wave fluxes are predominantly negative, which is primarily due to slightly negative pressure perturbations multiplied by large bulk outflow velocities along the $z$-axis after the outburst. The negative wave fluxes, however, play a minor role in the overall energy budgets (see later discussions). 
Between the bubbles and the forward shocks, sound waves of smaller perturbation amplitudes can be found. They are generated by a combination of physical processes including rarefaction waves behind the shock front, the RT and KH instabilities associated with the bubble disruption, and the interference of waves as they propagate. Some waves have lower amplitudes in the beginning, but could grow in amplitudes due to constructive interference (\citealt{Bambic19}). This effect could be seen from the horizontal profile of the pressure perturbations (Figure \ref{fig:delta_profile}; see the second peak behind the shock front). 

\begin{figure}
	\includegraphics[width=0.9\columnwidth]{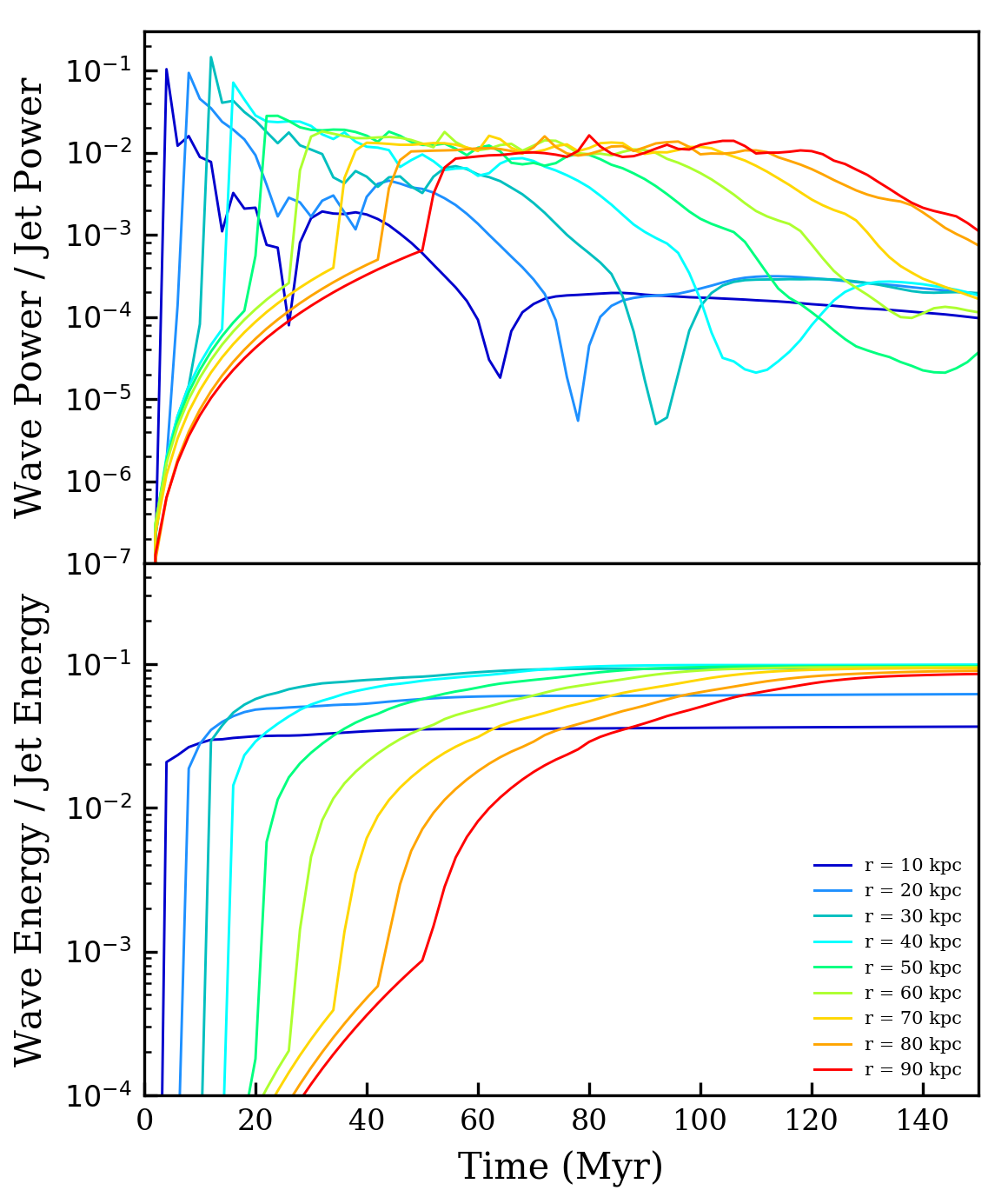}
	\centering
    \caption{Top: Evolution of compressional wave power normalized by the injected jet power measured at different radii. Bottom: Evolution of the cumulative compressional wave energy normalized by the injected jet energy.}
    \label{fig:acoustic_power_energy}
\end{figure}

In order to estimate how much jet energy could go into compressional waves, we use Eq.\ \ref{eq:wave_power} to measure the wave power crossing concentric spheres at different radii ranging from radius $r = 10$ kpc to $ 90$ kpc with equal spacing (shown in Figure \ref{fig:acoustic_power_energy} with different colors). The cumulative wave energy (normalized by the total injected energy) as a function of time measured at different radii is shown in the bottom panel. In an idealized case where the injection occurs from a point source at the center, there should be conservation of wave energies across different radii with a time delay when no explicit dissipation physics is included. However, the situation becomes more complex with the jet injection and subsequent bubble evolution. According to values shown in the bottom panel, we find that the integrated energies converge to $\sim 9 \%$ of total injected energy at late times of the simulation if the energies are measured beyond $r=30$ kpc. Within $r=30$ kpc, the measured energies are larger for larger radii. As discussed in \cite{Bambic19}, this is likely due to the fact that some of the waves are generated away from the cluster center, e.g., at the locations where bubbles are disrupted. Beyond $r=30$ kpc, we find that the cumulative wave energies measured at all radii converge to the same asymptotic value of $\sim 9\%$. This demonstrates that the sound-wave energy in our simulations is very well conserved, and the effect of numerical dissipation is negligible because the resolution within $r=100$ kpc in our simulation is maximally refined.

In our calculations for Figure \ref{fig:acoustic_power_energy}, we include only the outward-propagating waves, while in fact both outward and inward fluxes exist at the same time (see Figure \ref{fig:acoustic_flux_density}). In \cite{Bambic19}, they found the negative fluxes to be negligible in their simulations. We also compute the positive and negative power separately, and find that the contribution of negative fluxes to the power is about a hundred times lower than the contribution from positive fluxes. Therefore, the omission of negative power does not have a significant impact on our results.


In short summary, our single-jet simulation suggests that, for typical parameters suitable for the Perseus cluster, $\sim 9\%$ of the injected AGN jet energy could be transferred into compressional waves. This estimation should be regarded as an upper limit of the production efficiency of sound waves as it includes energies from both shocks and sound waves.  

\subsubsection{Contribution by shock waves}\label{Sec:shock}
\begin{figure}
	\includegraphics[width=\columnwidth]{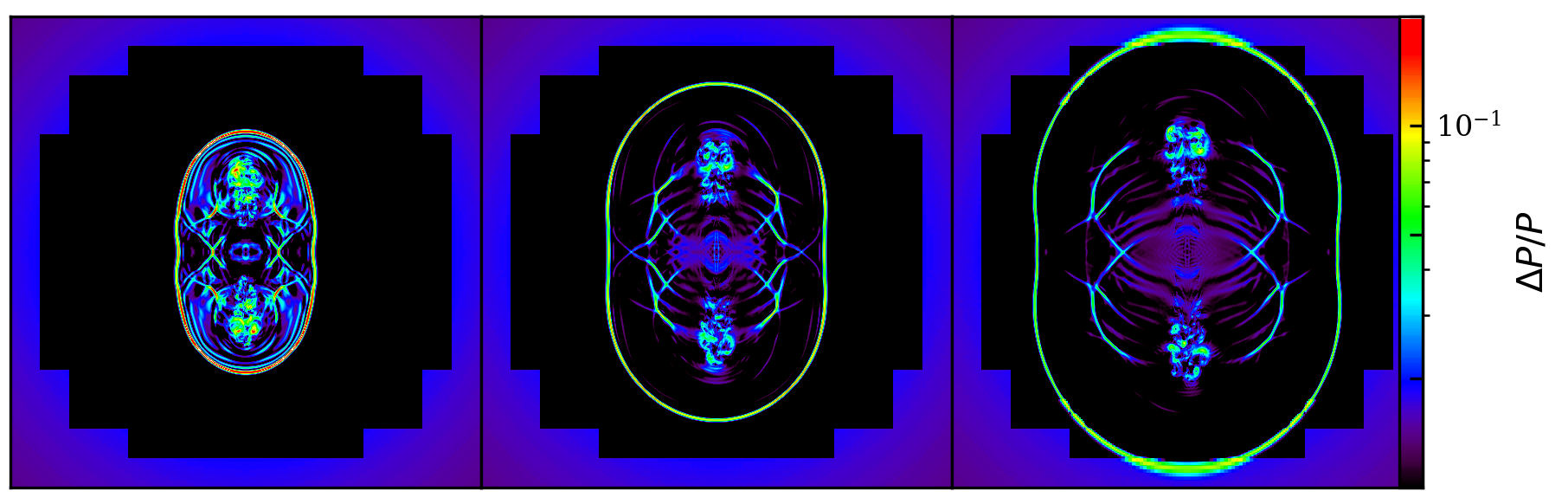}
	\centering
    \caption{Panels from left to right show slices of pressure jumps of neighboring cells, $\Delta P / P$, at $t=30, 50$, and 80 Myr. Two criteria are used in our analyses to identify shocks (in addition to negative velocity divergence), namely $\Delta P / P > 0.2$ and $\Delta P / P > 0.1$.}
    \label{fig:P_grad}
\end{figure}

Figure \ref{fig:P_grad} shows the pressure jumps across neighboring cells, $\Delta P/P$, at $t=30, 50$, and 80 Myr. In addition to negative velocity divergence, we apply two pressure-jump thresholds for identifying the shocks, $\Delta P/P > 0.2$ and $\Delta P/P > 0.1$. The former threshold is often adopted in previous numerical simulations \citep[e.g.,][]{Yang16b}, and the latter is used for comparison. Overall, the two selection criteria are effective in selecting the shocked cells, particularly at the outermost forward shocks. At later times, the pressure jumps at the forward shocks gradually decrease as the shocks dissipate their energies over time. They eventually become sound waves as indicated by the smaller pressure jump values than the thresholds. There are some grid artefacts at large radii in this figure due to the different refinement levels between neighboring blocks; however, note that the values in these regions are not included in our analyses.   


\begin{figure*}
	\includegraphics[width=0.8\textwidth]{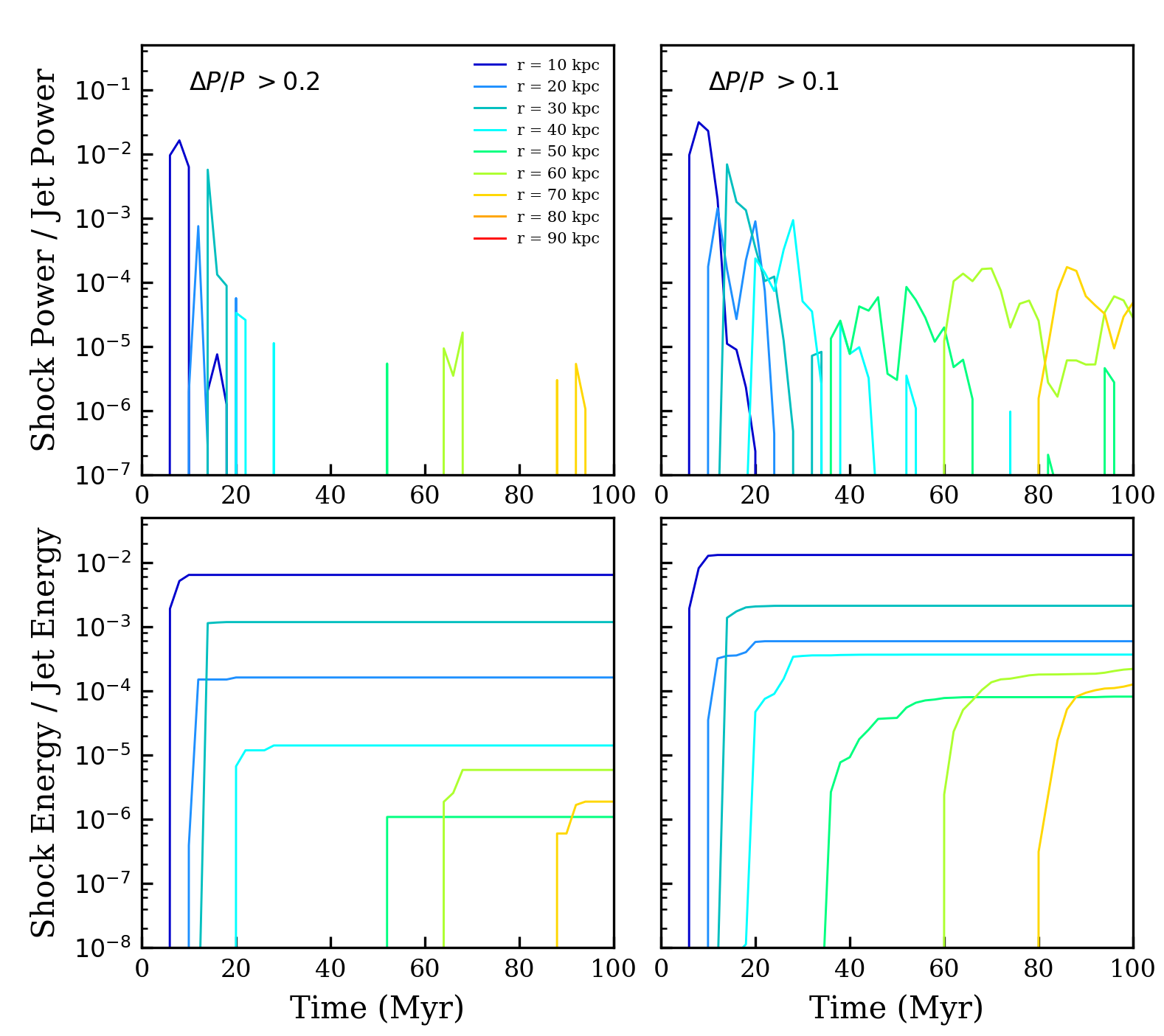}
	\centering
    \caption{Same as Figure \ref{fig:acoustic_power_energy} but only for grid cells containing shocks. Left and right columns show the results with the pressure-jump criterion of $\Delta P/ P > 0.2$ and $\Delta P/ P > 0.1$, respectively.}
    \label{fig:acoustic_power_energy_shock}
\end{figure*}

To quantify the impact by shock waves, we used the above criteria to select the shocked cells and compute their contribution to the wave power and energy (see Figure \ref{fig:acoustic_power_energy_shock}). The shock waves contribute the most wave power near the cluster center and their power drops very quickly as they propagate outward, especially for shocks with higher Mach numbers and greater pressure jumps. The same trends can also be seen in the cumulative wave energies in the bottom panel. For shocks with $\Delta P/P>0.2$, the shocks contribute to $\sim 0.7\%$ of the injected AGN energy, which is $\sim 23\%$ of the total wave energy at $r=10$ kpc (see bottom panel of Figure \ref{fig:acoustic_power_energy}). When a lower threshold of $\Delta P/P>0.1$ is adopted and more weaker shocks are included, $\sim 1.5\%$ of the injected energy goes into shocks, which takes $\sim 50\%$ of the total wave energy at $r=10$ kpc. However, the contributions by the shocks at larger radii are essentially negligible.     

\subsection{Self-regulated AGN feedback}\label{Sec:self_regulated}
In this section, we show results from the simulations including self-regulated AGN feedback. We vary three values of the feedback efficiency ($\epsilon=10^{-2}, 10^{-3}$, and $10^{-4}$) to investigate the impact of different levels of black hole activity on the production efficiencies of shocks and sound waves. The evolution of this set of simulations is similar to previous self-regulated feedback simulations \citep[e.g.,][]{Gaspari11, Yang16b, Li17}, in which a global thermal equilibrium can be established by balancing heating from AGN jets with radiative cooling within the cluster cores. In our simulations, the cluster initially radiatively cools and contracts for the first $t\sim 300$ Myr. Cold gas that is formed due to local thermal instabilities can be accreted onto the central SMBH and trigger an AGN outburst. The AGN jets heat the ICM via various processes including mixing, shocks, and turbulence. Eventually a self-regulated feedback cycle is established and the ICM properties reach a quasi-steady state after $t\sim 500$ Myr.  

\begin{figure}
	\includegraphics[width=\columnwidth]{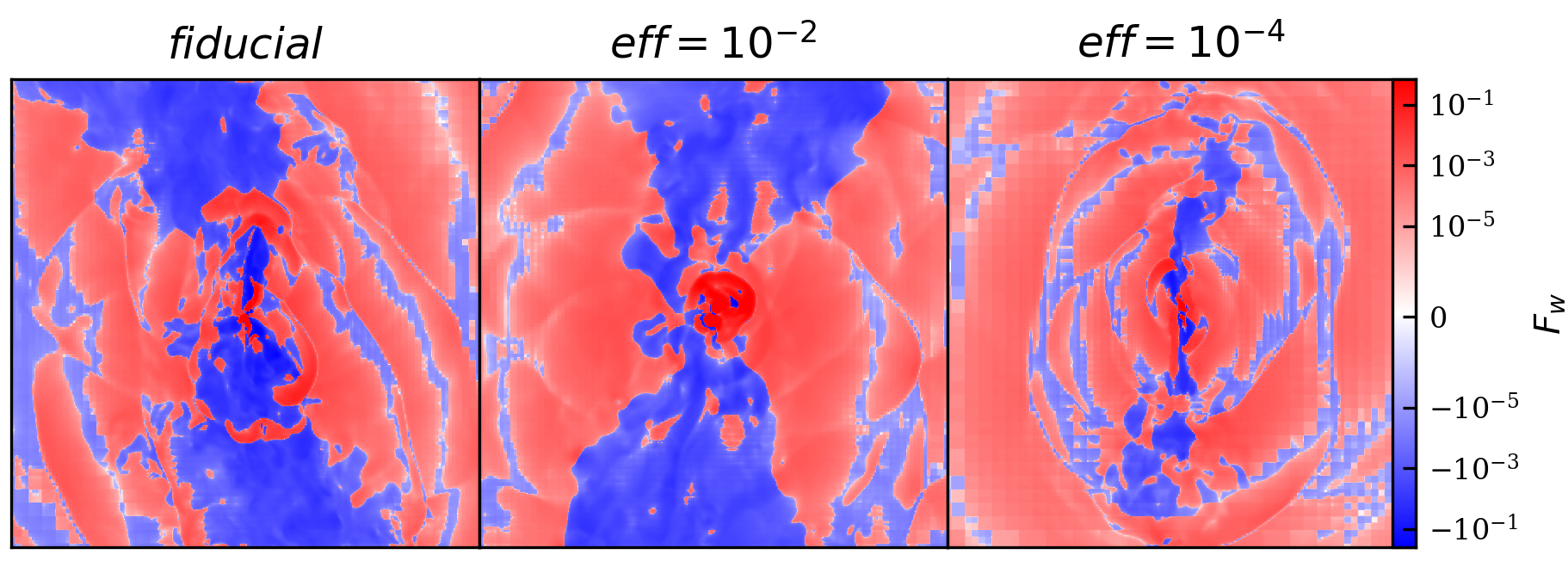}
    \caption{Slices of the flux density of compressional waves in the self-regulated feedback simulations for three different feedback efficiencies ($\epsilon=10^{-3}, 10^{-2}$, and $10^{-4}$). The region plotted in each panel is $270$ kpc on a side.}
    \label{fig:slice_self_regulated}
\end{figure}

Figure \ref{fig:slice_self_regulated} shows slices of compressional-wave flux density on the $y-z$ plane. Because in the self-regulated simulations the cluster profiles would evolve due to radiatively cooling during the initial $t\sim 300$ Myr, for computing the perturbed gas quantities in the subsequent analyses, we use the gas properties at $t=300$ Myr as the zero-point values (i.e., $P_0, \rho_0$). Same as Figure \ref{fig:acoustic_flux_density}, it clearly shows the outward propagation of waves from the inner region to outer radii, with negative flux densities associated with the jet-inflated bubbles (see discussion in \S~\ref{Sec:single_AGN}). The main difference between Figure \ref{fig:slice_self_regulated} and Figure \ref{fig:acoustic_flux_density} is the multiple ring structures due to the repeated AGN injections. Once the system reaches a quasi-equilibrium state, the overall distribution of waves appear to be qualitatively similar at different times and for different feedback efficiencies. 

\begin{figure*}
	\includegraphics[width=0.8\textwidth]{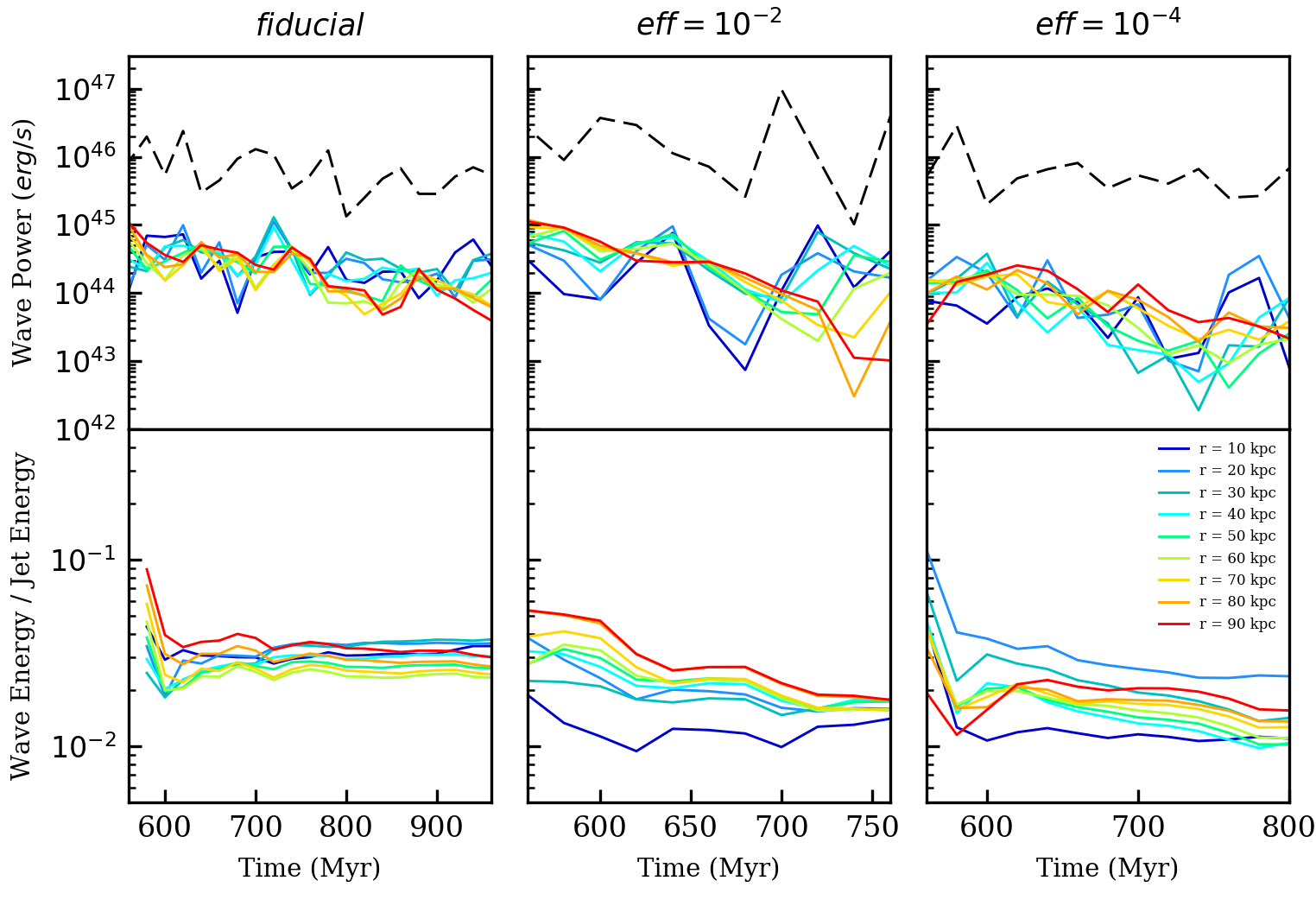}
	\centering
    \caption{Evolution of compressional wave power (top row) and cumulative energy (bottom row) for the self-regulated feedback simulations. The averaged AGN jet power is plotted as dashed lines. This plot shows the results of three simulations with different feedback efficiencies (fiducial case with $\epsilon=10^{-3}$ in the left panel, $\epsilon=10^{-2}$ in the middle panel, and $\epsilon=10^{-4}$ in the right panel). In all three simulations, the clusters reach a quasi-equilibrium state, and $\lesssim 3\%$ of the jet energy is converted into energy in compressional waves.}
    \label{fig:acou_e_self_regulated}
\end{figure*}

Again, we quantify the power and cumulative energy of compressional waves in the self-regulated feedback simulations and show the results in Figure \ref{fig:acou_e_self_regulated}. All the quantities are computed at each simulation output separated by a time interval of $\Delta t = 20$ Myr. Here we focus on the period of time when the systems are in a quasi-equilibrium state. The AGN jet power averaged every 20 Myr is overplotted in the dashed lines. One can see that the AGN jet power is more variable with a higher feedback efficiency ($\epsilon=10^{-2}$). On the other hand, the AGN jet power remains at a more constant level when the feedback efficiency is lower ($\epsilon=10^{-4}$). Despite the different patterns of AGN activity, the overall amplitudes of compressional wave power do not appear to vary significantly with feedback efficiencies. 


Because the instantaneous wave power as well as the AGN jet power show non-negligible variability, we evaluate the production efficiencies of compressional waves not based on the ratio between the instantaneous powers but the ratio between the cumulative wave energy and the cumulative jet energy (see the bottom panel of Figure \ref{fig:acou_e_self_regulated}). For the fiducial case ($\epsilon=10^{-3}$), we find that $\sim 3\%$ of the total injected energy goes into compressional waves for all radii. For the cases with high and low feedback efficiencies, the production efficiencies of compressional waves by AGN jets have somewhat larger variations across different radii, but the values are measured to be within the range of $\sim 1-3 \%$ as well. The efficiencies found here are a few times lower than the value for a single AGN outburst, and are much lower than the optimal values obtained by \cite{Bambic19}. The lower production efficiencies in the self-regulated feedback simulations could be due to the following factors. First, since the AGN injections are essentially random in time, the sound waves generated would be randomly phased, and therefore destructive interference could substantially weaken the waves. Second, each of the repeated AGN outburst generates shock waves whose perturbation amplitudes dominate over sound waves generated by previous events, and therefore the gas perturbations are dominated by the shocks driven by stronger AGN outbursts and their subsequent propagation and dissipation (as discussed below).   

\begin{figure*}
	\includegraphics[width=0.8\textwidth]{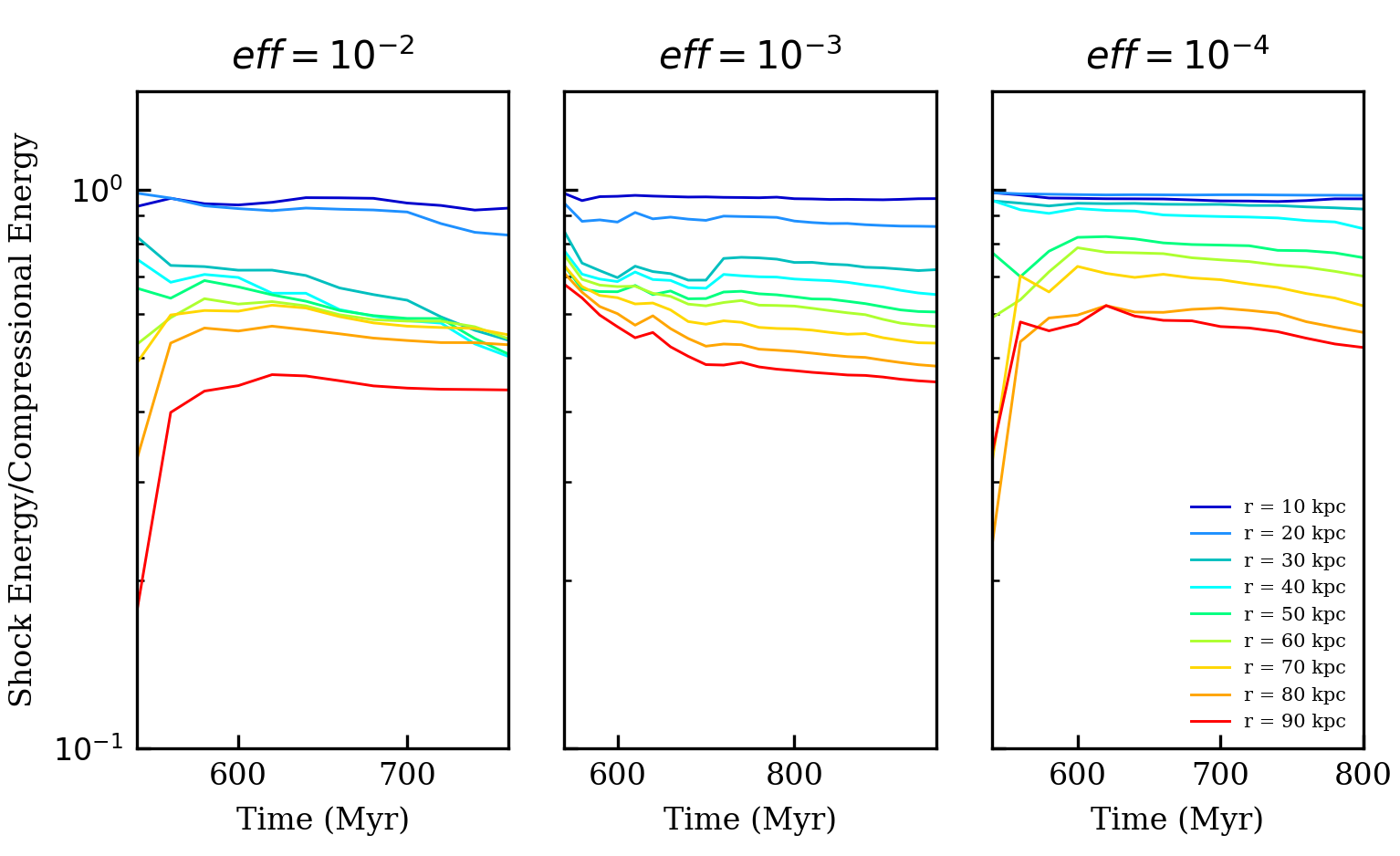}
	\centering
    \caption{The fractional contributions of shock waves to the total compressional wave energy as a function of time for the self-regulated feedback simulations. Panels from left to right show the simulations with different feedback efficiencies, $\epsilon=10^{-2}$, $\epsilon=10^{-3}$, and $\epsilon=10^{-4}$. Note that the pressure-jump criteria is set to be $\Delta P/ P > 0.1$ in this simulation. }
    \label{fig:shock_self_regulated}
\end{figure*}

The significant contribution of wave energies by shocks in the self-regulated feedback simulations could be seen in Figure \ref{fig:shock_self_regulated}, in which we show the ratios between the wave energies stored in shocks (with $\Delta P/P > 0.1$) and the total compressional wave energy (using the same method as in \S~\ref{Sec:shock}). For all feedback efficiencies, the fractional contribution of shocks is near 100\% in the inner 10-20 kpc. As the shocks propagate outward and dissipate, their contributions decrease with increasing radii. For simulations with $\epsilon=10^{-2}$, the power of AGN jets tends to be stronger, generating shocks with higher Mach numbers that dissipate more energy closer to the center. Therefore, the fractions of shock energies quickly drop within the inner $\sim 50$ kpc, and the shocks contribute to $\sim 40\%$ of the wave energies at $r=90$ kpc. In contrast, the decrease in the shock contribution with radius is more gradual for the other two cases, and $\sim 50\%$ of the wave energy is contributed by shocks at $r=90$ kpc. The values obtained here are much greater than those for the single-injection simulation, supporting our earlier claim that the compressional perturbations in the self-regulated feedback simulations are predominantly governed by the generation and dissipation of shock waves produced by the repeated AGN outbursts.

\section{Discussion}\label{Sec:discussion}

\subsection{Comparisons with previous works}\label{Sec:comparison}
Our work is along the lines of previous simulations that investigated the production efficiencies of sound waves driven by AGN outbursts. \cite{Tang17} studied energy injections from a point source in a uniform medium under spherical symmetry and found that $\lesssim 12\%$ of the injected energy can be carried away by sound waves. \cite{Bambic19} considered energy injections using bipolar jets in a stratified, isothermal atmosphere assuming axisymmetry and found that the sound wave production by jets is much more efficient, i.e., $\gtrsim 25\%$ of the jet energy can be converted into sound waves with optimal choices of parameters. They performed detailed parameter studies (in dimensionless quantities), and found that jets with relatively high velocities ($v_{\rm j}/c_{\rm s}=10^{1.75} - 10^{2.5}$, depending on density of the jet; $v_{\rm j}$ is the jet velocity and $c_{\rm s}$ is the sound speed of the medium; see their Figure 8) tend to transform over a quarter of the total jet energy into sound waves. When scaled to conditions in the Perseus cluster whose sound speed is $\sim 10^8$ cm s$^{-1}$, the jet velocities for optimal sound wave driving would correspond to $\sim 0.2-1c$, where $c$ is the speed of light. However, kpc-scale jets in the ICM are typically categorized as Faranoff-Riley (FR) I jets, suggesting that they have experienced significant interaction and deceleration with the ambient medium as they propagate to kpc scales \citep[e.g.,][]{Laing06}. Therefore, sub-relativistic jet velocities such as the value considered in our work may be more representative in typical conditions in CC clusters. According to the results in \cite{Bambic19}, our jet parameters ($v_{\rm j} \sim 10^{1.5}c_{\rm s}$, $\theta_{\rm j}=0^\circ$, and $\rho_{\rm j}\sim 0.1\rho_{\rm 0}$, where $\theta_{\rm j}$ is the opening angle of the jet, $\rho_{\rm j}=10^{-27}$ g cm$^{-3}$ is the jet density, and $\rho_0=10^{-26}$ g cm$^{-3}$ is the density scale used in \cite{Bambic19} for normalization) would produce sound waves at an efficiency of $\sim 0.07$. The production efficiency of $\lesssim 9\%$ from our single-jet simulation is thus in good agreement with the previous work of \cite{Bambic19}.  


Our results for the self-regulated feedback simulations could also be compared with previous simulations in Perseus-like clusters. Using a similar setup (with slightly different treatment of cold gas), \cite{Yang16b} used a different method to estimate the energy contribution from compressible (including shocks and waves) and incompressible perturbations (including gravity waves and turbulence). They found that the compressible modes only contribute to a few percent of the AGN energy (see their Figure 12), consistent with our current findings. \cite{Li17} also used simulations of self-regulated AGN feedback and they found that heating by weak shocks is an important mechanism for balancing radiative cooling in the ICM. Their simulations assume a feedback efficiency of $10^{-2}$, and they adopted a smaller threshold of $\Delta P/P=0.002$ for identifying shocks. Although we do not focus on the detailed AGN heating processes in this work, we computed the energy dissipation rates for the weak shocks identified in our simulations, and we verified that the results are comparable to Figure 1 in \cite{Li17} (not shown here). In addition, our analyses show that shocks could play an important role close to the cluster center, though their contribution decreases toward larger radii ($r>50$ kpc). This overall trend is consistent with the radial profiles of shock heating shown in Figure 5 of \cite{Li17}.   


\cite{Zhuravleva16} performed cross-spectrum analyses on the X-ray surface brightness fluctuations of the Perseus cluster and found that, within the inner 70 kpc region, $\sim 80\%$ of the total variance can be attributed to isobaric perturbations (e.g., gravity waves and turbulence) and less than $\sim 8\%$ of the variance to adiabatic fluctuations (e.g., shocks and sound waves). While this result may appear discrepant with the simulation study of \cite{Bambic19}, we find that, after including the cumulative effects of multiple AGN injections, the simulation results become more consistent with the few percent level contribution by the compressional waves found in the above observational study. This illustrates the importance of modeling self-regulated AGN feedback in the analyses of perturbations to be compared with the observational data.

    %
\subsection{Limitations of the simulations}\label{Sec:limitation}
This study is based on purely hydrodynamic simulations of momentum-driven AGN jets, and therefore some important physical mechanisms are not included. These processes may enhance or suppress the total output of compressional wave energy. We discuss the potential impact of each process below.

ICM transport processes such as viscosity and thermal conduction have been neglected in our simulations. Because of their attenuation effects, the production efficiencies of waves obtained in this work are likely to be upper limits. In \cite{Ruszkowski04a} and \cite{Ruszkowski04b}, they found that viscous dissipation of sound waves can be an important heating mechanism within cluster cores. Viscosity may also help to suppress the fluid instabilities at the surface of the AGN jet-inflated bubbles \citep{Reynolds05, Dong09, Kingsland19}, and thus may inhibit the production of sound waves near the surface of the bubbles. However, the coefficients of the transport processes in the ICM remain highly uncertain. Some observational studies indicate that the ICM viscosity must be suppressed compared to the standard Spitzer value \cite[e.g.,][]{Zuhone15, Su17, Zhuravleva19}. On the theoretical grounds, the transport coefficients must also be significantly suppressed so that the sound waves could propagate to large distances \citep{Fabian05, Zweibel17b}. Such suppression may be related to microscopic plasma instabilities in the magnetized, weakly collisional ICM \citep[e.g.,][]{Kunz11, Kingsland19, Drake21}. More works are demanded in order to understand the plasma properties of the ICM, which will give insights into the detailed dissipation processes of sound waves.  

Our simulations assume that the AGN jets are kinetic-energy dominated; however, the composition of AGN jets is uncertain \citep[e.g.,][]{Dunn04}. Previous works comparing jets of different compositions show that CR dominated jets tend to inflate bubbles that are more oblate and more stable against hydrodynamic instabilities \citep{Guo08, Yang19}. As a result, 
the generation of sound waves due to bubble disruption may be limited when CR jets are considered. Similarly, the inclusion of magnetic fields can also suppress the instabilities \citep{Ruszkowski08, Bambic18} and subsequent sound wave production. Owing to the above reasons, the production efficiencies of sound waves may be lower than values obtained in our simulations when CRs and magnetic fields are included. On the other hand, recent work by \cite{Kempski20} showed that streaming of CRs in the weakly collisional, magnetized ICM may induce instabilities and provide another mechanism for exciting sound waves. Therefore, future studies are demanded to determine the net effects of CRs on the production efficiencies of sound waves in the ICM.

Another factor that may affect our results is the resolution of our simulations. Because our simulations are performed in 3D, we could not afford to reach as high resolution as previous 2D simulations \citep{Bambic19}. Therefore, our simulations may not able to resolve sound waves on small scales and the production efficiencies may be somewhat underestimated. However, we note that, when the parameters are properly scaled to conditions of the Perseus cluster, our results are in fact comparable to the the findings of \cite{Bambic19} (see discussion in \S~\ref{Sec:comparison}). The agreement suggests that the resolution does not significantly influence our main conclusions. 



\section{Conclusions}\label{Sec:conclusion}
Sound wave dissipation is often considered as one of the important mechanisms for heating the ICM in CC clusters. Despite its importance, the production efficiency of sound waves by AGN jets as well as their dissipation remain unclear. In order to understand the energy budgets available for sound wave dissipation, we measured the amount of energy carried by compressional waves (including weak shocks and sound waves) using 3D hydrodynamic simulations in a Perseus-like cluster. Using simulations with a single injection and with self-regulated AGN feedback with varied feedback efficiencies, we quantified the fractions of jet energy that are converted into shocks and waves. Our main findings are summarized in the following. 

1. For the simulation of a single AGN outburst with typical parameters suitable for the Perseus cluster, we find that $\sim 9\%$ of the total injected energy from the central AGN can be transferred into compressional waves (Figure \ref{fig:acoustic_power_energy}). This amount of energy is consistent with previous 2D simulations of AGN jets for comparable parameters \citep{Bambic19}. 
Note that because we have neglected the transport processes of the ICM (e.g., thermal conduction and viscosity) and their dissipative effects on the waves, our estimates of the wave energies are likely to be upper limits.

2. For the self-regulated feedback simulations with feedback efficiencies ranging from $10^{-4}$ to $10^{-2}$, only $\sim 1-3\%$ of the total AGN jet energy is converted into compressional waves. This may be due to two reasons: (1) there is destructive interference among randomly phased waves from the repeated AGN injections, and (2) the shocks dominate the gas perturbations over sound waves generated by previous events and quickly dissipate their energies close to the cluster center.  

3. Shock waves in the single-jet case can have $\sim 20-30\%$ contribution to the total compressional wave energy in the inner radii, and their contribution drastically decreases as they propagate toward outer radii and dissipate. On the other hand, due to the frequent production of shocks generated by repeated AGN outbursts in the self-regulated case, the total compressional energy is almost completely dominated by shock waves in the inner 20 kpc, and the fractional contribution by shocks decreases to $\sim40-50\%$ at outer radii. 

Our results suggest that, when more realistic jet parameters and multiple AGN injections are considered, the production of sound waves is not as efficient as what was found in some of the previous simulations \citep{Tang17, Bambic19}, and thus sound wave dissipation may be a subdominant source of heating in CC clusters \citep{Hillel14}. We note, however, that our results are based on purely hydrodynamic simulations, and many important physical mechanisms are not included (see discussion in \S~\ref{Sec:discussion}). In particular, it remains an open question how well the magnetized, weakly collisional ICM could be modeled as a hydrodynamic fluid. Also, the transport coefficients of the ICM are still highly uncertain. Future simulations incorporating these important microphysical mechanisms, together with observational constraints on the transport coefficients from measurements resolving the scales of the collisional mean free paths, will provide valuable insights into the plasma properties of the ICM and the detailed processes of sound wave dissipation.

\section*{Acknowledgements}

SCW and HYKY acknowledge support from Yushan Scholar Program of the Ministry of Education of Taiwan and Ministry of Science and Technology of Taiwan (MOST 109-2112-M-007-037-MY3). The simulations are performed and analyzed using computing facilities operated by the National Center for High-performance Computing and the Center for Informatics and Computation in Astronomy at National Tsing Hua University. FLASH was developed largely by the DOE-supported ASC/Alliances Center for Astrophysical Thermonuclear Flashes at University of Chicago. Data analysis presented in this paper was conducted with the publicly available {\it yt} visualization software \cite{Turk11}.

\section*{Data Availability}

The data underlying this article will be shared on reasonable request to the corresponding author.



\bibliographystyle{mnras}
\bibliography{mnras} 








\bsp	
\label{lastpage}
\end{document}